\documentclass[showpacs,amssymb,eqsecnum,aps]{revtex4}
\usepackage{epsf}
\usepackage{color}
\newcommand{\be}{\begin{equation}}
\newcommand{\ee}{\end{equation}}
\newcommand{\fracd}[2]{\displaystyle\frac{#1}{#2}}

\newcommand{\de}{\delta\varepsilon}

\newcommand{\nnbb}{\nonumber\\}

\begin{document}

\title{Thermal one- and two-graviton Green's functions in the temporal gauge}

\author{F. T. Brandt, B. Cuadros-Melgar and F. R. Machado}
\affiliation{Instituto de F\'{\i}sica,
Universidade de S\~ao Paulo,
S\~ao Paulo, SP 05315-970, BRAZIL}

\date{\today \\ }

\begin{abstract}
The thermal one- and two-graviton Green's function are computed
using a temporal gauge. In order to handle the extra poles which are
present in the propagator, we employ an ambiguity-free
technique in the imaginary-time formalism. For temperatures $T$ high
compared with the external momentum, we obtain the  leading $T^4$ as well
as the subleading $T^2$ and $\log(T)$ contributions to the 
graviton self-energy. The gauge fixing independence of the leading
$T^4$ terms as well as the Ward identity relating the
self-energy with the one-point function are explicitly verified. 
We also verify the \'{}t Hooft identities for the subleading
$T^2$ terms and show that the logarithmic part
has the same structure as the residue of the ultraviolet pole of the
zero temperature graviton self-energy. We explicitly compute the
extra terms generated by the {\it prescription poles} and verify that
they do not change the behavior of the leading and sub-leading
contributions from the {\it hard thermal loop} region.
We discuss the modification of the solutions of the dispersion
relations in the graviton plasma induced by the subleading $T^2$ contributions.
\end{abstract}


\maketitle

\section{Introduction}
One of the main motivations for the first attempts to compute
the self-energy at finite temperature was the study of dispersion
relations of a graviton plasma and the related interesting phenomena of
anti-damping and wave propagation
\cite{gross:1982cvkikuchi:1984npgribosky:1989yk,Rebhan:1991yr}. 
For temperatures $T$ high compared with the external momentum, but well
below the Planck scale, the complete tensor structure of the leading
one-loop contributions,  proportional to $T^4$, was calculated for the
first time in Ref. \cite{Rebhan:1991yr}. Later some subleading
contributions of order $T^2$ were computed, including the
contributions of thermal scalar matter and radiation \cite{deAlmeida:1994wy},
and subsequently all terms proportional to $T^4$, $T^2$ and $\log(T)$
where computed taking into account thermal loops of gravitons
\cite{Brandt:1998hd}. 

When the internal graviton lines are included, the gauge dependence
which arises from the choice of gauge fixing in the gravitational
action becomes an issue. In Ref. \cite{Brandt:1998hd} the graviton
self-energy was computed employing the Feynman-de Donder
gauge with an arbitrary gauge fixing parameter. While the subleading
contributions are gauge dependent, the leading $T^4$ 
contributions to the self-energy as well as to the one-point function
are gauge fixing independent and satisfy the Ward identity. This last
property is also true for the contributions from matter and radiation,
being consistent with a gauge invariant effective action for hard
thermal loops interacting with gravity.

One can go further into the question of gauge dependence 
by considering a class of non-covariant gauges of the kind that has
been employed in gravity at zero temperature \cite{Matsuki:1979rt,Delbourgo:1981dx,Delbourgo:1981dt,Capper:1982rc,Capper:1982vk,Matsuki:1985xc}. 
At finite temperature non-covariant temporal gauges would be even
more appropriate, since the Lorentz covariance is already broken by the heat
bath but the rotational invariance is preserved.
Despite the other well known advantages of the temporal gauge, finite
temperature calculations have been performed only in 
Yang-Mills theories both in imaginary and in the real time formalisms
\cite{Kajantie:1985xx,Kobes:1989cr,Kobes:1989up,James:1990itJames:1990fdJames:1991dz,Leibbrandt:1994ki,Leibbrandt:1996cu,Brandt:1999nf}.
This can be partially understood in view of the complexity of the
gravitational interaction and so
explicit calculations in non-covariant gauges have been restricted to
the zero temperature case.
Another reason for the lack of popularity of the temporal gauge in
gravity is that, in contrast with the situation in Yang-Mills theory,
the zero temperature graviton self-energy is not transverse
\cite{Delbourgo:1981dt,Capper:1982rc}. However, this should not be a
very important concern in the finite temperature case where the
transversality property is expected to be violated in general.
A more important difficulty in the temporal gauge is the problem of
spurious singularities arising from the $n=0$ terms in the Matsubara
sums \cite{kapusta:book89lebellac:book96das:book97}, which is even more 
severe in the case of gravity in view of the higher powers of $n$
in the denominator of the temporal gauge graviton propagator. This
situation was improved after the development of an ambiguity-free
technique to perform perturbative calculations at finite temperature in
the temporal gauge \cite{Leibbrandt:1994ki,Leibbrandt:1996cu}.
Originally this technique was tested using zeta-functions
to compute the Matsubara sums and later it
was applied to the calculation of the gluon self-energy using
the standard method of introducing thermal distributions
by replacing the Matsubara frequency sum with a contour integral in
the complex plane of the zero component
of the internal momentum \cite{Brandt:1999nf}.

The purpose of the present work is to apply the 
Leibbrandt's prescription to the calculation of the thermal
one- and two-graviton Green's functions in a class of temporal gauges.
We will show explicitly how this approach leads to a well defined
result which can be expressed in terms of forward scattering
amplitudes of thermal gravitons \cite{Frenkel:1992tsBrandt:1997se}
plus contributions from prescription poles. We will also show how the
ghost interactions effectively decouple leaving only thermal gravitons
in the forward scattering amplitudes. We provide the explicit results
for the leading and sub-leading hard thermal loop contributions and
show that the prescription poles do not change the hard thermal loop
behavior.

In section {\bf II} we will present the Lagrangian for the
graviton field and the corresponding Feynman rules in a class of
temporal gauges. We will also illustrate the basic approach with
the simplest one-loop calculation, namely the one-graviton
function (tadpole). In section {\bf III} we describe how the thermal
graviton self-energy can be split in two parts. The first part arises
from the on-shell poles of thermal graviton, and it is expressed in
terms of forward scattering amplitudes, while the second part is
generated by poles in the complex energy plane which are
characteristic of the temporal gauge prescription. We obtain from the
forward scattering amplitudes the leading $T^4$ and the subleading
$T^2$ and ${\log} T$ contributions.
In section {\bf IV} we explicitly calculate the contributions from
the prescription poles and compare the results with the high
temperature limit of the forward scattering expression.
In section {\bf V} we employ the hard thermal loop results,
up to the subleading $T^2$ contributions, to investigate the
modification of the solutions of the dispersion relations in a
gravitational plasma. We will discuss our results in section {\bf VI}.  

\section{Lagrangian, Feynman rules and basic definitions}
The graviton field, $\phi_{\mu\nu}$, can be defined as a small
perturbation around the flat space-time metric, $\eta_{\mu\nu}$, as follows 
\be\label{gravdef}
g_{\mu\nu}(x) = \eta_{\mu\nu} + \kappa\,\phi_{\mu\nu}(x),
\;\;\;\; \kappa^2 = 32\,\pi\,G.
\ee
Here $G$ is Newton's constant and $g_{\mu\nu}$ is the metric
tensor. The Einstein Lagrangian density is given by
\be
{\cal L} = \frac{2}{\kappa ^2} \sqrt{-g}\, g^{\mu\nu}\, R_{\mu\nu}
\ee
where $R_{\mu\nu}$ is the Ricci tensor given by
\begin{eqnarray}
R_{\mu\nu} & = & \partial_\nu\,\Gamma^{\alpha}_{\mu\alpha} -
                 \partial_\alpha\,\Gamma^{\alpha}_{\mu\nu} -
                 \Gamma^{\alpha}_{\mu\nu}\,\Gamma^{\beta}_{\alpha\beta} +
                 \Gamma^{\alpha}_{\mu\beta}\,\Gamma^{\beta}_{\nu\alpha}\nnbb  
\Gamma^{\alpha}_{\mu\nu} & = & \frac 1 2 g^{\alpha\beta}
\left(\partial_\mu\,g_{\beta\nu} + \partial_\nu\,g_{\beta\mu}
- \partial_\beta\,g_{\mu\nu}\right)
\end{eqnarray}
It is clear from the previous expressions that the Einstein
Lagrangian is an infinity series in powers of $\kappa$
(an infinity number of terms arises both from the inverse metric
$g^{\mu\nu}$ and from the determinant $g$).
Each power $\kappa^n$ will come out multiplied by a combination of 
tensor scalar products of $n$ tensor fields $\phi$ and two derivatives
$\partial \phi$. Performing a systematic expansion in powers of
the coupling constant $\kappa$, it is straightforward to obtain
the tree-level Feynman rules corresponding to the terms which are
quadratic, cubic, etc \footnote{The derivation of the momentum space
Feynman rules as well as the subsequent calculations of diagrams 
were all performed using the mapleV3 version of the computer algebra
package HIP \cite{hsieh:1992ti}.}. Before we show the explicit form of
these vertices, let us recall that the invariance of the
Einstein action under general coordinate
transformations (gauge transformations) imply the existence of
Ward identities relating all the vertices down to the quadratic term
(see the appendix \ref{apa}). The identity given by Eq. (\ref{transv1})
shows explicitly the usual problem of inverting the free quadratic part of a
gauge invariant Lagrangian. Following the standard procedure of introducing
a gauge fixing condition and ghost fields, 
we add the following two terms to the Einstein Lagrangian 
\cite{leibbrandt:1987ki}
\be\label{fix}
{\cal L}_{fix}=-{1\over 2 \alpha}\eta^{\mu\nu}(n^{\rho}
                           \phi_{\rho\mu})(n^{\sigma}\phi_{\sigma\nu})
\ee
and
\be\label{ghost1}
{\cal L}_{ghost} = -n^{\mu}\chi^{\nu}{\delta\phi_{\mu\nu}\over
  \de^{\lambda}}\eta^{\lambda},
\ee
where $\chi^{\mu}$ and $\eta^{\mu}$ are the gravitational
Faddeev-Popov ghost vector fields, $n^\mu$ is the axial vector and
$\alpha$ is a constant gauge parameter.
Using Eq. (\ref{delphi}), we obtain the following explicitly form for
the ghost Lagrangian
\be\label{ghost}
{\cal L}_{ghost} = \chi^{\nu}\left[
n_\lambda\,\partial_\nu + \eta_{\lambda\nu}\,n \cdot\partial
+\kappa\left(n^\mu\,\phi_{\mu\lambda}\partial_\nu 
+\phi_{\nu\lambda}\, n \cdot\partial
+ n^\mu\,(\partial_\lambda\phi_{\mu\nu}) \right)\right]\eta^{\lambda}.
\ee
Notice that, unlike Yang-Mills theory, ghosts remain coupled to
the gravitons even for the choice $\alpha=0$. However, our explicit  
calculation will show that the decoupling occurs when the loop
integrations are performed.

We have now all the basic ingredients to perform perturbative
calculations in thermal gravity. The graviton propagator
can now be obtained inverting the quadratic term of 
${\cal L} + {\cal L}_{fix}$. Our choice of gauge fixing is such that
even the bare graviton propagator is already dependent on fourteen
independent tensors as shown in table 1 (we will employ this
same basis in order to obtain the tensor structure of the thermal self-energy)
\footnote{These tensors, which have been employed in 
Ref. \cite{Rebhan:1991yr} in thermal gravity, were introduced in
Ref. \cite{Matsuki:1979rt} in the context of gravity at zero
temperature formulated in the axial gauge. In the present paper we employ the
same sorting of Ref. \cite{Rebhan:1991yr}, but we normalize the tensors
in such a way that they are all dimensionless.}.
Using this tensor basis it is possible to obtain the following
compact form for the graviton propagator
\begin{eqnarray}\label{gravprop}
{\cal D}_{\lambda \beta,\, \rho \sigma}(k) =&&{{1}\over{  (k^2 +
    i \epsilon)}} \left\{ I^1 _{\lambda \beta,\, \rho \sigma} - {1 \over
    D-2} I^2 _{\lambda \beta,\, \rho \sigma} + \alpha {k^4\over
       n_0^2\,({k\cdot u})^2} 
\left[{\cal T}^8_{\lambda \beta,\, \rho \sigma}+
{k^2\over(k\cdot u)^2}{\cal T}^{12}_{\lambda \beta,\,\rho\sigma} 
- {\cal T}^{11}_{\lambda \beta,\, \rho \sigma}\right]\right\} \, , 
\end{eqnarray}
where
\begin{eqnarray}
&& I^1 _{\mu \nu ,\,\rho \sigma} = {1 \over 4} (d_{\mu \kappa} d_{\nu\lambda} + 
d_{\mu \lambda} d_{\nu \kappa})(d_{\rho \kappa} d_{\sigma \lambda} + d_{\rho \lambda} d_{\sigma \kappa}) ,\nonumber \\
&& I^2 _{\mu \nu ,\,\rho \sigma} = d_{\mu \kappa} d_{\nu \kappa} d_{\rho
  \lambda} d_{\sigma \lambda}; \quad d_{\mu \nu} = \eta _{\mu \nu}
-{{k_\mu u_\nu } \over {k\cdot u}} \, ,\nonumber 
\end{eqnarray}
are convenient linear combinations of the tensors in Table I.
As we can see the graviton propagator has the usual poles at $k^2=0$
as well as the poles at $k\cdot u = k_0 =0$.

\begin{table}[htbp]\label{tab1}
\begin{center}
\begin{tabular}{c l}\hline
${\cal T}^1 _{\mu \nu ,\, \rho \sigma}=$&$  \eta _{\mu \rho} \eta_{\nu \sigma} + \eta _{\mu \sigma} \eta_{\nu \rho}$ \\ & \\
${\cal T}^2 _{\mu \nu ,\, \rho \sigma}=$&$ \eta_{\mu \rho} u_\nu u_\sigma +
\eta_{\mu \sigma} u_\nu u_\rho + \eta_{\nu \rho} u_\mu u_\sigma +
\eta_{\nu \sigma} u_\mu u_\rho$ \\ & \\ 
${\cal T}^{3} _{\mu \nu ,\,\rho \sigma} =$&$  u_\mu u_\nu u_\rho  u_\sigma$
 \\ & \\
${\cal T}^4 _{\mu \nu ,\,\rho \sigma} =$&$ \eta_{\mu \nu} \eta_{\rho\sigma}$
\\ & \\
${\cal T}^5 _{\mu \nu ,\,\rho \sigma} =$&$ \eta_{\mu \nu} u_\rho u_\sigma
+ \eta_{\rho \sigma} u_\mu u_\nu$ \\ & \\
${\cal T}^6 _{\mu \nu ,\,\rho \sigma} =$&$ \fracd{1}{k\cdot u}
\left[(\eta_{\mu \rho} k_\nu
+\eta_{\nu \rho} k_\mu) u_\sigma + (\eta_{\mu \sigma} k_\nu +
\eta_{\nu \sigma} k_\mu) u_\rho +\right. $\\ 
$$&$\left. \;\;\;\;+ (\eta_{\mu \rho} u_\nu +\eta_{\nu \rho} u_\mu) k_\sigma +
(\eta_{\mu \sigma} u_\nu + \eta_{\nu \sigma} u_\mu) k_\rho\right]$\\ & \\
${\cal T}^{7} _{\mu \nu ,\,\rho \sigma} =$&$ 
\fracd{1}{k\cdot u}\left(k_\mu u_\nu u_\rho u_\sigma +
k_\nu u_\mu u_\rho u_\sigma + k_\rho u_\mu u_\nu u_\sigma + k_\sigma
u_\mu u_\nu u_\rho\right) $\\ & \\
${\cal T}^8 _{\mu \nu ,\,\rho \sigma} =$&$ \fracd{1}{k^2}\left(
\eta_{\mu \rho} k_\nu k_\sigma
+\eta_{\mu \sigma} k_\nu k_\rho + \eta_{\nu \rho} k_\mu k_\sigma +
\eta_{\nu \sigma} k_\mu k_\rho\right) $ \\ & \\
${\cal T}^{9} _{\mu \nu ,\,\rho \sigma} =$&$ \fracd{1}{k^2}\left(
k_\mu k_\nu u_\rho u_\sigma
+ k_\rho k_\sigma u_\mu u_\nu \right)$ \\ & \\
${\cal T}^{10} _{\mu \nu ,\,\rho \sigma} =$&$ \fracd{1}{(k\cdot u)^2}
\left[(k_\mu u_\nu + k_\nu u_\mu)(k_\rho u_\sigma + k_\sigma u_\rho)\right]$ \\ & \\
${\cal T}^{11} _{\mu \nu ,\,\rho \sigma} =$&$ \fracd{1}{k^2\,k\cdot u}\left(
u_\mu k_\nu k_\rho k_\sigma +
u_\nu k_\mu k_\rho k_\sigma + u_\rho k_\mu k_\nu k_\sigma + u_\sigma
k_\mu k_\nu k_\rho\right) $ \\ & \\
${\cal T}^{12} _{\mu \nu ,\,\rho \sigma} =$&$  \fracd{1}{k^4}
k_\mu k_\nu k_\rho k_\sigma  $ \\ & \\
${\cal T}^{13} _{\mu \nu ,\,\rho \sigma} =$&$  \fracd{1}{k^2}\left(
\eta_{\mu \nu} k_\rho k_\sigma 
+ \eta _{\rho \sigma} k_\mu k_\nu \right)$\\ & \\
${\cal T}^{14} _{\mu \nu ,\, \rho \sigma}= $&$  \fracd{1}{k\cdot u}\left[
  \eta_{\mu\nu}\left(k_\rho u_\sigma + u_\rho k_\sigma\right) 
+ \eta_{\rho\sigma}\left(k_\mu u_\nu + u_\mu k_\nu\right)\right] $ \\ 
 \\ \hline
\end{tabular}\caption{The fourteen independent tensors built from
$\eta_{\mu\nu}$, $k_\mu$ and $u_\mu\equiv n_\mu/n_0$
and satisfying the symmetry conditions
${\cal T}^{i}_{\mu\nu,\rho\sigma} = {\cal T}^{i}_{\nu\mu,\rho\sigma}
= {\cal T}^{i}_{\mu\nu,\sigma\rho} = {\cal T}^{i}_{\rho\sigma,\mu\nu}$.}
\end{center}
\end{table}

The first and second order terms in $\kappa$ yield the following 
three and four graviton vertices respectively
\begin{eqnarray}\label{vert3}
 V^3_{\alpha\beta,\,\rho\lambda,\,\delta\gamma}(k_1,k_2,k_3)&=&{\kappa\over
  4}\left\{ \left[ k_2\cdot
  k_3(\eta_{\alpha\beta}(\eta_{\rho\lambda}\eta_{\delta\gamma}-\eta_{\rho\delta}\eta_{\lambda\gamma})+
  4\eta_{\alpha\delta}(\eta_{\beta\rho}\eta_{\gamma\lambda}-\eta_{\rho\lambda}
  \eta_{\beta\gamma}))\right.\right.\nnbb
&&+2{k_2}_\alpha({k_3}_\beta(\eta_{\lambda\gamma}\eta_{\rho\delta}-\eta_{\rho\lambda}\eta_{\delta\gamma})+2{k_3}_\rho(\eta_{\beta\lambda}\eta_{\delta\gamma}-2\eta_{\beta\delta}\eta_{\lambda\gamma}))\nnbb 
&&+2{k_2}_\rho(2
  {k_3}_\alpha\eta_{\beta\lambda}\eta_{\delta\gamma}+
  {k_3}_\lambda(2\eta_{\beta\gamma}\eta_{\alpha\delta}-\eta_{\alpha\beta}\eta_{\delta\gamma}))+2{k_2}_\delta{k_3}_\rho\nnbb
&&\left.\times
(\eta_{\alpha\beta}\eta_{\lambda\gamma}-2\eta_{\beta\lambda}
\eta_{\alpha\gamma})\right]+symmet.\; 
  on \; (\alpha\leftrightarrow \beta),\,(\rho\leftrightarrow
  \lambda),\nnbb
&&,\left.(\delta\leftrightarrow \gamma)\right\}+permut.\; of\; (k_1,\alpha,\beta),(k_2,\rho,\lambda),(k_3,\delta,\gamma),
\end{eqnarray}
\begin{eqnarray}\label{vert4}
\hspace*{-2cm}&
V^4_{\alpha\beta,\,\rho\lambda,\,\delta\gamma,\,\tau\sigma}(k_1,k_2,k_3,k_4)
&\hspace*{-.2cm}={\kappa^2\over
  16}\left\{\left[k_3\cdot
 k_4((\eta_{ \alpha\beta}\eta_{ 
\rho\lambda} - 2\eta_{ \alpha\rho}
\eta_{ \beta\lambda}) 
 (\eta_{ \delta\gamma}\eta_{\tau\sigma} - \eta_{
 \delta\tau}\eta_{\gamma\sigma})+ 8 (\eta_{ \alpha\delta}\eta_{ \beta\rho}\right. \right.\nnbb
&&+ \eta_{ \alpha\rho}\eta_{ 
\beta\delta}-\eta_{ \alpha\beta}\eta_{
 \delta\rho})(\eta_{\lambda\gamma}\eta_{\tau\sigma}-\eta_{\sigma\gamma}\eta_{\tau\lambda}) + 8\eta_{\rho\tau}\eta_{\alpha\delta}(\eta_{ \beta\gamma}\eta_{\sigma \lambda} - \eta_{
  \beta\sigma}\eta_{ \gamma\lambda}))\nnbb
&&+ 4 {k_3}_\alpha 
  (2 {k_4}_\rho\eta_{ \beta\lambda} - {k_4}_\beta\eta_{ \rho\lambda})(\eta_{ \delta\gamma}\eta_{\tau\sigma}- \eta_{ \delta\tau}\eta_{ \gamma\sigma}) + 16( {k_3}_\rho{k_4}_ 
\alpha\eta_{ \beta\delta} - {k_3}_\alpha {k_4}_\beta\eta_{
  \delta\rho})
\nnbb
&&\times(\eta_{\gamma\sigma}\eta_{ \lambda\tau} - \eta_{
  \gamma\lambda}\eta_{\tau\sigma})+ 8({k_3}_\alpha
{k_4}_\delta
 + {k_3}_\delta{k_4}_\alpha)  
(\eta_{ \rho\lambda}\eta_{ \beta\gamma} - 2\eta_{ \gamma\lambda}\eta_{ 
\beta\rho})\eta_{\tau\sigma} \nnbb
 && +16 {k_3}_\alpha{k_4}_\delta(
\eta_{ \rho\tau}(2\eta_{ \beta\sigma}\eta_{ \gamma\lambda} - \eta_{ \beta \gamma}\eta_{\sigma\lambda}) 
  + \eta_{ \gamma\sigma}(2\eta_{ \rho\tau}\eta_{ \beta\lambda} -
  \eta_{ \rho\lambda}\eta_{ \beta\tau}))\nnbb
 && -
  16 {k_3}_\delta{k_4}_\alpha\eta_{\rho\tau}\eta_{\beta\gamma}\eta_{\sigma\lambda} 
+ 2({k_3}_\tau {k_4}_\delta\eta_{ \gamma \sigma} - {k_3}_\delta {k_4}_\gamma\eta_{\tau\sigma}) 
  (\eta_{ \alpha\beta}\eta_{ \rho\lambda} - 2\eta_{ \alpha\rho}\eta_{ 
\beta\lambda}) \nnbb 
&&\left.+ 8 ({k_3}_\tau{k_4}_\delta\eta_{ \gamma\lambda}- {k_3}_\delta{k_4}_\gamma\eta_{ \lambda\tau})(2\eta_{ \beta\sigma}\eta_{ \alpha\rho}- \eta_{ \rho\sigma}\eta_{ 
\alpha\beta})\right]\nnbb
&&\left.+symmet.\;  on \;(\alpha\leftrightarrow
  \beta),\,(\rho\leftrightarrow \lambda),\,(\delta\leftrightarrow
  \gamma),\,(\tau\leftrightarrow \sigma)\right\}\nnbb
&&+permut.\;of \;(k_1,\alpha,\beta),(k_2,\rho,\lambda),
(k_3,\delta,\gamma),(k_4,\tau\sigma)\, .
\end{eqnarray}
We have verified that these vertices are in agreement with the Ward
identities described in the Appendix \ref{apa}. 

Finally, the quadratic and the interacting term in the ghost
Lagrangian (\ref{ghost}) yields the ghost propagator 
\be\label{ghprop}
{\cal D}^{\rm ghost}_{\lambda \mu} (k) = 
  {i} \left[ {1 \over {2(n.k)^2}} k_\lambda n_\mu 
- {1 \over {n.k}} \eta _{\lambda \mu} \right] 
\ee
and the graviton-ghost-ghost vertex
\be\label{Ggg}
V^{Ggg} _{\mu \kappa,\, \rho \nu} ({k_1},{k_2},{k_3}) = i 
\kappa(\eta_{\rho \mu} \eta_{\nu \kappa} n.{k_2} + 
       \eta_{\rho \kappa} n_\mu {k_2}_\nu
     + \eta_{\nu  \kappa} n_\mu {k_1} _\rho ) 
+ \mu\leftrightarrow\kappa
\, ,
\ee
respectively.

\subsection{The one-point function}\label{tadsec}
In order to introduce our notation and the basic method of calculation
we will rederive here the result for the thermal one-point function.
The one-point function is interesting by itself, since it is directly
related to the energy momentum tensor derived from the effective
action \cite{Rebhan:1991yr}. It also provides the simplest 
non-trivial example of a one-loop calculation in gravity. Indeed, 
in contrast with the zero temperature case, the finite temperature
one-point function is non-zero, being exactly proportional to $T^4$. 
For that reason it will play an important r\^ole in the Ward
identities obeyed by the hard thermal loop Green's functions.

The relevant diagrams are shown in the Fig. I.
\begin{figure}[t!]
\begin{center}
  \hspace{.8\textwidth} \vbox{ \epsfxsize=0.3\textwidth
    \epsfbox{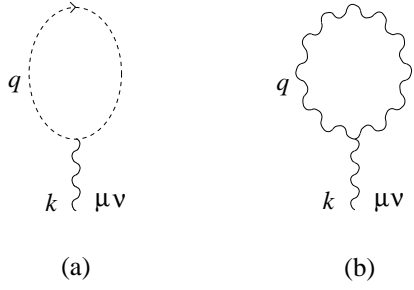}}
\caption{Diagrams contributing to the one-graviton function in the 
one-loop approximation. The curly lines represent gravitons and the
dashed lines represent ghosts.}       
\end{center}
\end{figure}
Using the imaginary time formalism
\cite{kapusta:book89lebellac:book96das:book97} the
Eqs. (\ref{ghprop}) and (\ref{Ggg}) give the following
contribution for the ghost loop diagram shown in figure 1(a)
\be\label{oneghost}
\Gamma_{\mu\nu}^{\rm ghost} = \kappa\,T\,\sum_{q_0} 
\int\frac{d^{D-1}\,\vec q}{(2\pi)^{D-1}}\; \eta_{\mu\nu};\;\;\;
q_0 = 2\pi\,i\,n\, T;\;\;\; n=0,\pm 1,\pm 2, \dots.
\ee

\noindent
Throughout this work the Matsubara sums like that one in 
Eq. (\ref{oneghost}) will be computed using the standard and elegant 
relation \cite{kapusta:book89lebellac:book96das:book97}
\be\label{sumq0}
T\sum_{q_0} f(q_0) = \int_{-i\infty+\delta}^{i\infty+\delta}
\frac{{\rm d} q_0}{2\pi i}\left[f(q_0)+f(-q_0)\right]
\frac 1 2 \coth(\frac{q_0}{2T}) =
\int_{-i\infty+\delta}^{i\infty+\delta}
\frac{{\rm d} q_0}{2\pi i}\left[f(q_0)+f(-q_0)\right]
\left(\frac 1 2 + \frac{1}{{\rm e}^{\frac{q_0}{T}} -1}\right).
\ee
In general, the vacuum part of the amplitudes
(terms which arise from the factor $1/2$ inside 
the round bracket of Eq. (\ref{sumq0})) may be divergent in the
limit $D\rightarrow 4$ and so the arbitrary dimension $D$ 
provides a regulator for the vacuum piece of the thermal Green's
functions as usual \cite{'tHooft:1972fiBollini:1972ui}.

The tadpole diagram provides the simplest example of an 
effective decoupling of the ghost graviton interaction in the temporal
gauge (this is not a trivial property at non-zero
temperature). Indeed, substituting (\ref{sumq0}) into
Eq. (\ref{oneghost}) we can see that the vacuum piece vanishes as a
consequence of the identity
\be\label{Dreg}
\int {d^{D-1}\,\vec q}\; |\vec q|^{r} = 0.
\ee
The thermal piece also vanishes since we can close the contour in the
right hand side of the $q_0$ plane without enfolding any poles.

The contribution from the graviton loop in figure 1(b) is a
little bit more involved. After some straightforward tensor algebra we
obtain from Eqs. (\ref{gravprop}) and (\ref{vert3}) the following result
\be\label{tadgrav}
\Gamma_{\mu\nu} = \kappa\,T\,\sum_{q_0} 
\int\frac{d^{D-1}\,\vec q}{(2\pi)^{D-1}}\frac{D}{8}
\left[2(D-3)\frac{q_\mu\,q_\nu}{q^2} - (D-5)\,\eta_{\mu\nu}\right].
\ee

\noindent
It is interesting to notice that the gauge parameter $\alpha$
from the graviton propagator has already canceled out at the
integrand level.

Let us now compute Eq. (\ref{tadgrav}) with the help of
formula (\ref{sumq0}). As in the case of the ghost loop diagram, 
the contribution proportional to $\eta_{\mu\nu}$ vanishes. The
dimensional regularized vacuum piece will also vanish and we are left
with only the following expression
\be\label{tadgrav2}
\Gamma_{\mu\nu} = 2\,\kappa\int_{-i\infty+\delta}^{i\infty+\delta}
\frac{{\rm d}\, q_0}{2\pi\, i}\frac{1}{{\rm e}^{\frac{q_0}{T}}-1}
\int\frac{{\rm d}^{3}\,\vec{q}}{(2\pi)^{3}}
\frac{q_\mu\,q_\nu}{q^2}.
\ee
Closing the contour in the right hand side plane the pole at
$q_0=|\vec q|$ gives the following contribution
\be\label{tadgrav3}
\Gamma_{\mu\nu} = -\kappa\left.\int_{0}^{\infty}
\frac{{\rm d}|\vec q|}{(2\pi)^{3}}
\frac{|\vec q|^3}{{\rm e}^{\frac{|\vec q|}{T}}-1}
\int {\rm d}\Omega{\hat q_\mu\,\hat q_\nu}\right|_{q_0=|\vec q|},
\ee
where we have introduced $\hat q_\mu = q_\mu/|\vec q|$ and 
$\int {\rm d}\Omega$ is the integration over all 
directions of $\vec q$. Finally, using the formula \cite{gradshteyn}
\be\label{intq}
\int_0^{\infty}\frac{x^{\nu -1}}
{{\rm e}^{{{x}/{T}}}-1}{\rm d}x
= \Gamma(\nu)\,\zeta(\nu)\,T^\nu
\ee
we obtain
\be\label{tadgrav4}
\Gamma_{\mu\nu} = -\kappa\, \frac{\pi^2\, T^4}{30}
\left.\int \frac{{\rm d}\Omega}{4\,\pi}\,
{\hat q_\mu\hat q_\nu}\right|_{q_0=|\vec q|} = 
 -\kappa\, \frac{\pi^2\, T^4}{90}
\left(4\,u_\mu u_\nu - \eta_{\mu\nu}\right),
\ee

\noindent
where we have employed the quantity $u\equiv (1,0,0,0)$, which
coincides with the vector representing the local rest frame of the
plasma and was introduced in the table I.

\section{Thermal forward scattering contributions to the graviton self-energy}
The diagrams which contribute to the graviton self-energy are shown
in figure 2. The relevant Feynman rules for the propagators and vertices are
all given in the previous section. Let us first consider the ghost
loop diagram shown in figure 2(a). As we can see from the structure
of the ghost propagator in Eq. (\ref{ghprop}) the integrand
will involve a combination of fractions of the following type
\be\label{ghfrac}
\frac{1}{(q\cdot u)^m\,[(k+q)\cdot u]^n},\;\;\; m,n=0,1,2 .
\ee
\begin{figure}[t!]
\begin{center}
  \hspace{0.8\textwidth} \vbox{ \epsfxsize=0.65\textwidth
    \epsfbox{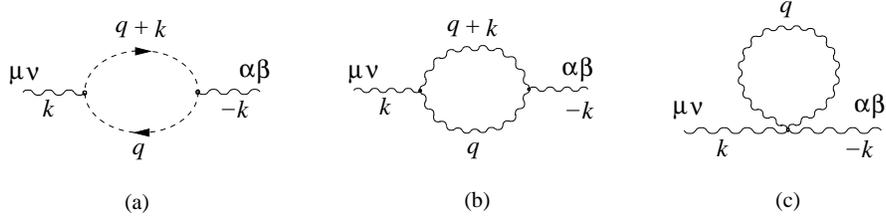}}
\caption{Diagrams contributing to the graviton self-energy.
The curly lines represent gravitons and the 
dashed lines represent ghosts. The external momentum $k$ is inward.}       
\end{center}
\end{figure}
Before trying to perform the loop momentum integrations explicitly
it is convenient to simplify the integrand using well known 
algebraic identities and change of variables which may reduce the
number of terms considerably. Indeed, we have found that using a partial
fraction decomposition of the quantities shown in Eq. (\ref{ghfrac})
and a shift $q\rightarrow q-k$ in the resulting partial fractions
containing powers of $\left[(q+k)\cdot u\right]^{-1}$,
leads to the simplest possible result given by
\begin{eqnarray}\label{ghloop}\small
\hspace*{-1.5 cm}\Pi^{gh}_{\mu\nu,\alpha\beta}&=&\sum_{q_0}\int \frac{dq^{D-1}}{(2\pi)^{D-1}}\left[\frac{k^2 k \cdot q\, u_\mu u_\nu u_\alpha u_\beta}
{k\cdot u^3 q\cdot u}-\frac{1}{2}\frac{k^2 u_\mu u_\nu q_\alpha u_\beta  }{ k\cdot u
 ^2 q\cdot u}+\frac{1}{2}\frac {u_\mu k_\nu u_\alpha q_\beta }{k\cdot u \, q\cdot u}-\frac{1}{2}
\frac {k \cdot q\, u_\mu k_\nu u_\alpha  u_\beta}{k\cdot u^2 q\cdot u}\right.\nonumber \\
\hspace*{-1.5 cm}&-&\frac{1}{2}\frac {u_\mu q_\nu u_\alpha q_\beta}{q\cdot u^2}+\frac{1}{2}\frac {u_\mu q_\nu k_\alpha u_\beta}{k\cdot u q\cdot u}
+\frac{1}{2}\eta_{\alpha\mu}\eta_{\beta\nu}+\frac{1}{2}\frac{q_\mu u_\nu k_\alpha u_\beta }
{k\cdot u \, q \cdot u}-\frac{1}{2}\frac {k \cdot q\, u_\mu u_\nu k_\alpha u_\beta}{k\cdot u^2\, 
q\cdot u}+\frac{1}{2}\frac{k \cdot q\, q_\mu u_\nu u_\alpha u_\beta}{k\cdot u\, q\cdot u^2}\nonumber \\
\hspace*{-1.5 cm}&-&\frac{1}{2}\frac{k^2 
 u_\mu q_\nu u_\alpha u_\beta}{k\cdot u^{2}\,q\cdot u}+\frac{1}{2}\frac {k \cdot q\, u_\mu q_\nu u_\alpha u_\beta}{k\cdot u\, q\cdot u^2}-\frac{1}{2}\frac{q_\mu u_\nu q_\alpha u_\beta}{q\cdot u^{2}}-\frac{1}{2}\frac {q_\mu u_\nu u_\alpha q_\beta}{q\cdot u^2}-\frac{1}{2}\frac {u_\mu q_\nu q_\alpha u_\beta}
{q\cdot u^2}+\frac{1}{2}\eta_{\alpha\nu}\eta_{\beta\mu}\nonumber \\
\hspace*{-1.5 cm}&-&\frac{1}{2}\frac {k^2 q_\mu u_\nu u_\alpha u_\beta}{k\cdot u^2\, q\cdot u}+\frac{1}{2}\frac {u_\mu k_\nu q_\alpha u_\beta}
{ k\cdot u\,q\cdot u}+\frac{1}{2}\frac{k_\mu u_\nu u_\alpha q_\beta}
{ k\cdot u \, q\cdot u}+\frac{1}{2}
\frac {k_\mu u_\nu q_\alpha u_\beta}{k\cdot u\, q\cdot u}-\frac{1}{2}\frac{k \cdot q\, 
k_\mu u_\nu u_\alpha u_\beta }{k\cdot u^{2}q\cdot u}+\frac{1}{2}\frac {u_\mu q_\nu u_\alpha k_\beta}{k\cdot u\, q\cdot u}\nonumber \\
\hspace*{-1.5 cm}&-&\frac{1}{2}\frac {k \cdot q\, u_\mu u_\nu u_\alpha k_\beta}{k\cdot u^2\, q\cdot u}+\frac{1}{2}\frac {q_\mu u_\nu u_\alpha k_\beta}{k\cdot u\, q\cdot u}-
\frac{1}{2}\frac{k^2 u_\mu u_\nu u_\alpha q_\beta}{k\cdot u^2 \, q\cdot u}+\frac{1}{2}
\frac {k \cdot q\, u_\mu u_\nu u_\alpha q_\beta}
{k\cdot u \, q\cdot u^2}-\frac{1}{2}
\frac {k\cdot q^{2} u_\mu u_\nu u_\alpha u_\beta}{k\cdot u^{2}\, q\cdot u^{2}}\nonumber \\
\hspace*{-1.5 cm}&+&\left.\frac{1}{2}
\frac{k \cdot q\, u_\mu u_\nu q_\alpha u_\beta}{ k\cdot u\, q\cdot  u^2}
\right]
\end{eqnarray}\normalsize
\begin{center}
\end{center}
This procedure (partial fractions and then shifts)
has been employed previously in the case of the Yang-Mills
theory \cite{Brandt:1999nf}. In contrast with the present 
thermal gravity result given by Eq. (\ref{ghloop}),
the axial gauge Yang-Mills ghost loop vanishes at the
integrand level. Notice that the partial fraction decomposition is
justified since the integrands are regularized accordingly.

Let us now consider the diagrams shown in figures 2(b) and 2(c). 
An important difference between these diagrams and the ghost loop is
that while the ghost loop contains {\it only} the  poles at $q_0=0$,
the structure of the graviton propagator in 
Eq. (\ref{gravprop}) is such that there are {\it also} 
the usual simple poles in the right hand
side plane located at $q_0=|\vec q|$ and $q_0=|\vec q+\vec k| - k_0$
for the diagram in figure 2(b) and at $q_0=|\vec q|$ for the diagram
in figure 2(c)
(notice that $k_0$ is an imaginary quantity at this stage of the calculation).
In order to use the contour method of integration 
described in section \ref{tadsec}, we will employ the following
prescription for the poles at $q_0=0$ \cite{Leibbrandt:1994ki}
\be\label{presc}
{1\over q_0^r}\rightarrow\lim_{\mu\rightarrow 0}
{q_0^r\over (q_0^2-\mu^2)^r}\, .
\ee
With this prescription the temporal gauge poles are moved away from the
imaginary axis and we are allowed to employ the formula (\ref{sumq0}).
The $q_0$ integral can then be performed closing the contour 
of integration in the right hand side of the $q_0$ plane, as we did
in the previous section in the case of the one point function.
The contributions from the {\it prescription poles} 
located at $q_0=\mu$ will be analyzed in the next section.

We now follow the steps explained in the Appendix A of
Ref. \cite{Brandt:1999nf}. Basically this consists in 
the use of Eq. (\ref{sumq0}) taking into account only the contributions
from the poles located on the right hand side plane at
$q_0=|\vec q|$ and $q_0=|\vec q+\vec k| - k_0$. Then, in the residues from
the poles at $|\vec q+\vec k| - k_0$ we perform the shift
$\vec q\rightarrow \vec q - \vec k$ and use the property
$\coth(x+k_0) = \coth(x)$. This yields the following expression in
terms of thermal {\it forward scattering amplitudes}
\be\label{gravForw}
\left. \Pi_{\mu\nu,\alpha\beta} \right |_{FS} =-\frac{1}{(2\pi)^3}
\int\frac{d^3 q}{2|\vec q|}\frac{1}{{\rm e}^{\frac{|\vec q|}{T}}-1}\frac 1 2
\left\{
  \begin{array}{c}
 \epsfbox{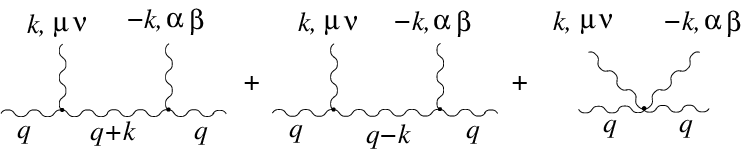}
  \end{array}
+q\leftrightarrow -q
\right\}_{q_0=|\vec q|},
\ee
where the factor $1/2$ in front of the curly brackets takes into
account the symmetry of the graphs in Figs. 2(b) and 2(c). It is
understood that the external graviton lines with momentum $q$ 
are contracted with the tensor given by the curly bracket of
Eq. (\ref{gravprop}).

We remark that the gauge parameter dependence of Eq. (\ref{gravForw})
involves only linear terms in $\alpha$. This can be understood since
the quadratic powers of $\alpha$ which could in principle arise from the
propagator in Eq. (\ref{gravprop}) do not have the on-shell poles.
Another interesting property of Eq. (\ref{gravForw}) is that it does
not involve {\it thermal ghosts}.

The forward scattering expression in Eq. (\ref{gravForw}) is 
very convenient when considering the {\it hard thermal loop}
contributions which arise from the region where the internal momentum
$q$ is of the order of the temperature $T$, which is large compared
to the external momentum $k$. In this regime we can expand the
denominators in Eq. (\ref{gravForw}) as follows
\be\label{exp1}
{1\over k^2 \pm 2\,k\cdot q} = \pm {1\over 2\,k\cdot q} -
{k^2\over (2\,k\cdot q)^2} + \cdots .
\ee
The leading hard thermal loop contribution is obtained by considering all the
integrands which are of degree two in the internal momenta $q$. After
some straightforward but very tedious algebra we were able
to express the leading contribution in the following rather compact form
\be
\label{twopoint}
\left.\Pi_{\mu \nu , \alpha \beta}^{{\rm lead}}\right|_{{\rm FS}}= 
-\kappa^2\, \frac{\pi^2\,T^4}{30}
\int\frac{\text{d}\Omega }{4\pi }\frac 1 2\,
\left[\left(k \cdot \frac{\partial}{\partial\hat q}\right)
\frac{\hat q_\mu \hat q_\nu \hat q_\alpha \hat q_\beta }
{\hat q\cdot k} - \eta_{\mu\alpha}\,\hat q_\nu \hat q_\beta
                - \eta_{\nu\alpha}\,\hat q_\mu \hat q_\beta
                - \eta_{\mu\beta}\,\hat q_\nu \hat q_\alpha
                - \eta_{\nu\beta}\,\hat q_\mu \hat q_\alpha
\right]_{q_0=|\vec q|}, 
\ee
where we have employed the formula (\ref{intq})
and $\hat q$ have the same meaning as in the Eq. (\ref{tadgrav4}).

One can easily verify that this leading $T^4$ contribution is related
to the one-graviton function in Eq.  (\ref{tadgrav4}) by the Ward
identity in Eq. (\ref{v2v1}) (this result is also in agreement with
the calculations performed in the Feynman-de Donder
gauge \cite{Rebhan:1991yr,Brandt:1998hd}). Since we expect that the
leading $T^4$ contributions are generated by a gauge independent
effective action, the contributions from the prescription poles
in Eq. (\ref{presc}) should not modify the leading $T^4$ behavior. 
This will be confirmed by our explicit calculation in the next section.

Let us now consider the subleading contributions which are generated
when we expand the integrand of Eq. (\ref{gravForw}) up to terms of
degree zero in $q$. By power counting these will be of order $T^2$.
In order to obtain the full tensor structure generated by the 
expression (\ref{gravForw}) it is convenient to use the following tensor 
decomposition
\be\label{selfgen}
\Pi^{\mu\nu\alpha\beta}=\sum_{l=1}^{14}C_l {\cal T}_l^{\mu\nu,\alpha\beta},
\ee
where the tensors ${\cal T}_l^{\mu\nu,\alpha\beta}$  are given in
Table I. The coefficients $C_l$ are obtained solving the system of
14 equations 
\be\label{sys1}
\sum_{l=1}^{14}\left({\cal T}_i^{\mu\nu,\alpha\beta}
                {{\cal T}_l}_{\mu\nu,\alpha\beta}\right)\,C_l=
                \Pi_i,\;\;\; i=1,2,\dots,14,
\ee 
where the quantities $\Pi_i$ are the following 
projections of the graviton self-energy
\be\label{proj}
\Pi_i=\Pi_{\mu\nu, \alpha\beta}\, {\cal T}_i^{\mu\nu, \alpha\beta},
\;\;\; i=1,...,14.
\ee
Each one of these projections can be expanded using Eq. (\ref{exp1}).
The integrals over the modulus of $\vec q$ can be easily performed
using Eq. (\ref{intq}) (they yield the $T^2$ factor)
and the angular integrals are all straightforward. Inserting
the results for $\Pi_i$ into Eq. (\ref{sys1}) and solving for $C_l$,
we obtain
\begin{eqnarray}\label{ct2}
C^{T^2}_1&=&\left[{k^4 L(k)\over \vec{k}^2}\left({1\over 12}+{5\over
  192}{k^2\over \vec{k}^2}\right)+{1\over
  36}\vec{k}^2-{5\over 576}{k^4\over \vec{k}^2}-{1\over
  144}k^2-{1\over 30}\frac{\alpha}{n_0^2} k^4\right]\kappa^2T^2 \nnbb
C^{T^2}_2&=&\left[{k^6 L(k)\over \vec{k}^4}\left({7\over 32}+{25\over
  192}{k^2\over \vec{k}^2}\right)-{25\over 576}{k^6\over \vec{k}^4}
-{1\over 18}{k^4\over \vec{k}^2}-{1\over
  36}k^2+{4\over 45}\frac{\alpha}{n_0^2} k^4\right]\kappa^2T^2 \nnbb
C^{T^2}_3&=&\left[{k^8 L(k)\over \vec{k}^6}\left({15\over 16}+{175\over
  192}{k^2\over \vec{k}^2}\right)-{175\over
  576}{k^8\over\vec{k}^6}-{55\over
  288}{k^6\over\vec{k}^4}+{1\over 18}{k^4\over \vec{k}^2}+{1\over
  9}k^2-{4\over 15}\frac{\alpha}{n_0^2} k^4\right]\kappa^2T^2 \nnbb
C^{T^2}_4&=&\left[\left(-{1\over 16}+{5\over
  192}{k^2\over \vec{k}^2}\right){k^4 L(k)\over \vec{k}^2}-{4\over
  15}\frac{\alpha}{n_0^2}(\vec{k}^4+k^2\vec{k}^2-{1\over 12}k^4)-{5\over 576}{k^4\over
  \vec{k}^2}+{5\over 288}k^2\right]\kappa^2T^2 \nnbb
C^{T^2}_5&=&\left[{25\over 192}{k^8 L(k)\over \vec{k}^6}+{1\over
  18}\vec{k}^2-{25\over 576}{k^6\over \vec{k}^4}+{5\over 288}{k^4\over
  \vec{k}^2}
+{1\over 9}k^2-{1 \over 15}\frac{\alpha}{n_0^2}(4k^2\vec{k}^2+{14\over 3}k^4)\right]\kappa^2T^2
  \nnbb
C^{T^2}_6&=&k_0^2\left[-\left({7\over 32}+{25\over
  192}{k^2\over \vec{k}^2}\right){k^4 L(k)\over \vec{k}^4}+{25\over 576}{k^4\over \vec{k}^4}
+{1\over 18}{k^2\over \vec{k}^2}-{4\over 45}\frac{\alpha}{n_0^2} k^2\right]\kappa^2T^2
  \nnbb
C^{T^2}_7&=&k_0^2\left[-\left({15\over 16}+{175\over
  192}{k^2\over \vec{k}^2}\right){k^6 L(k)\over \vec{k}^6}+{175\over 576}{k^6\over
  \vec{k}^6}+{55\over 288}{k^4\over \vec{k}^4}
-{1\over 18}{k^2\over \vec{k}^2}-{1\over 18}+
 {4\over 15}\frac{\alpha}{n_0^2} k^2\right]\kappa^2T^2
  \nnbb
C^{T^2}_8&=&k^2\,\left[\left({13\over 96}+{31\over
  96}{k^2\over \vec{k}^2}+{25\over 192}{k^4\over \vec{k}^4}\right)
{k^2 L(k)\over \vec{k}^2}-{25\over 576}{k^4\over \vec{k}^4}
-{13\over 144}{k^2\over \vec{k}^2}-{1\over 48}+\left({4\over
  45}\vec{k}^2+{11\over 90}k^2\right)\frac{\alpha}{n_0^2}\right]\kappa^2T^2
  \nnbb
C^{T^2}_9&=&k^2\,\left[\left({15\over 16}+{55\over 32}{k^2\over \vec{k}^2}+{175\over
  192}{k^4\over \vec{k}^4}\right)
{k^4 L(k)\over \vec{k}^4}-{175\over 576}{k^6\over \vec{k}^6}-{65\over
  144}{k^4\over \vec{k}^4}-{11\over 72}{k^2\over \vec{k}^2}+{2\over
  45}\frac{\alpha}{n_0^2} k^2\right]\kappa^2T^2
  \nnbb
C^{T^2}_{10}&=&k_0^2\,\left[\left({23\over 32}+{55\over 32}{k^2\over \vec{k}^2}+{175\over
  192}{k^4\over \vec{k}^4}\right)
{k^4 L(k)\over \vec{k}^4}-{175\over 576}{k^6\over \vec{k}^6}-{65\over
  144}{k^4\over \vec{k}^4}-{23\over 288}{k^2\over \vec{k}^2}+{1\over
  12} \right.\nnbb
&&\left. -{4\over 15}\frac{\alpha}{n_0^2}(\vec{k}^2+{4\over 3}k^2)\right]\kappa^2T^2  \nnbb
C^{T^2}_{11}&=&{k^2\,k_0^2\over \vec{k}^2}\left[-\left({1\over 2}+{35\over
  24}{k^2\over \vec{k}^2}+{175\over 192}{k^4\over \vec{k}^4}\right)
{k^2 L(k)\over \vec{k}^2}+{175\over 576}{k^4\over \vec{k}^4}
+{35\over 96}{k^2\over \vec{k}^2}+{1\over 24}+{2\over 15}\frac{\alpha}{n_0^2}\vec{k}^2\right]\kappa^2T^2
  \nnbb
C^{T^2}_{12}&=&{k^4\over \vec{k}^2}\left[\left({1\over 6}+{29\over
  24}{k^2\over \vec{k}^2}+{95\over 48}{k^4\over \vec{k}^4}+{175\over
  192}{k^6\over \vec{k}^6}\right)L(k)-{175\over 576}{k^4\over \vec{k}^4}
-{155\over 288}{k^2\over \vec{k}^2}-{5\over 24}-{2\over 45}\frac{\alpha}{n_0^2}\vec{k}^2\right]\kappa^2T^2
  \nnbb
C^{T^2}_{13}&=&k^2\,\left[\left({1\over 16}+{5\over
  48}{k^2\over \vec{k}^2}+{25\over 192}{k^4\over \vec{k}^4}\right)
{k^2 L(k)\over \vec{k}^2}-{25\over 576}{k^4\over \vec{k}^4}
-{5\over 288}{k^2\over \vec{k}^2}-\left({14\over
  45}\vec{k}^2+{1\over 3}k^2\right)\frac{\alpha}{n_0^2}\right]\kappa^2T^2
  \nnbb
C^{T^2}_{14}&=&k_0^2\,\left[-{25\over 192}{k^6 L(k)\over \vec{k}^6}+{25\over 576}{k^4\over \vec{k}^4}
-{5\over 288}{k^2\over \vec{k}^2}-{1\over 18}+{1\over
  15}\frac{\alpha}{n_0^2}(4\vec{k}^2+{14\over 3}k^2)\right]\kappa^2T^2
  \nnbb
\end{eqnarray}
where
\be\label{Llog}
L(k)={k_0\over 2|\vec{k}|}{\log} {k_0+|\vec{k}|\over k_0-|\vec{k}|}-1 \, .
\ee

There are some properties of the $T^2$ contributions which are worth
stressing. First, the $T^2$ contributions show their gauge dependence
explicitly through the gauge parameter $\alpha$.
Each of these gauge parameter dependent terms have two powers of
momentum relative to the corresponding $\alpha$-independent ones
(the correct mass dimension is provided by $n_0^2$ in the denominators). 
Secondly, the simple Ward identity satisfied by the leading
$T^4$ contributions is no longer true for the sub-leading
contributions. In the Appendix \ref{apb} we derive the more general
\'{}t Hooft  identities and we verify that the following identity
is satisfied 
\be\label{hooft}
\chi^{(0)}_{\mu\nu\lambda}\Pi^{\mu\nu\alpha\beta}_{T^2}=-\chi^{(1)\,T^2}_{\mu\nu\lambda}V^{2\, \mu\nu\alpha\beta}
\ee
where $\chi^{(1)}_{\mu\nu\lambda}$ is represented by the diagram shown
in Fig. 3. The $T^2$ contribution to $\chi^{(1)}_{\mu\nu\lambda}$ is
computed in detail in the Appendix B.

\begin{figure}[h!] 
\hspace{-0.4\textwidth}\begin{center}
     \hspace{.8\textwidth} \vbox{ \epsfxsize=0.25\textwidth
    \epsfbox{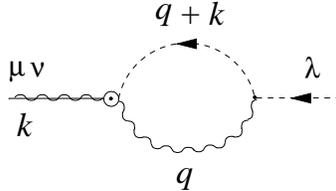}}
\caption{ The source-ghost diagram. The full/wave
   line on the left represents the external source.}
\end{center}
\end{figure}

We have proceed even further with the hard thermal loop expansion of
Eq. (\ref{gravForw}) and computed the contributions
from the integrands of degree minus 2 in $q$. After integration
these yield the ${\log} T$ terms. 
We have verified that the ${\log} T$ contributions of all the projections
$\Pi_i$, (see Eq. (\ref{proj})) are simply related to the corresponding
projections of the ultraviolet divergent part of zero temperature
graviton self-energy. The zero temperature results were computed using
the gauge choice $\alpha=0$ in \cite{Capper:1982rc}. Setting
$\alpha=0$ in our general result we have verified that
\be\label{logT}
\left.\Pi^{\log}_{\mu\nu,\alpha\beta}\right|_{\rm FS} =
{\log}(T)\, \Pi^{\epsilon}_{\mu\nu,\alpha\beta},
\ee
where $\Pi^{\epsilon}_{\mu\nu,\alpha\beta}$ is the residue of the
ultraviolet divergent zero temperature contribution computed in
$D=4-2\epsilon$ dimensions.
The verification of this property in the case of gravity formulated in
the temporal gauge complements similar results obtained
in the Feynman-de Donder gauge \cite{Brandt:1998hd}
as well as in the case of the Yang-Mills theory
\cite{Brandt:1999gm,Brandt:1999nf}.
Since our calculation has been performed for arbitrary values of
$\alpha$, we present complete results in the Appendix C.

\section{The contributions from prescription poles}
Let us now consider the terms which arise
from the poles located at $q_0=\mu$, where $\mu$ is 
the quantity introduced in Eq. (\ref{presc}).
It is convenient to express these contributions directly in terms of
the projections defined by Eq. (\ref{proj}). Each one of the fourteen
projections can be expressed as follows
\be\label{projpresc}
\Pi_{i}^{{\rm presc}}= \lim_{\mu\rightarrow 0}
\sum_{r=1}^{4}\sum_{q_0}\int
d^{D-1}\vec{q}\left[
{f_i^r(k_0,\vec{k}\cdot\vec{q},\vec{q}^{\,2},\vec{k}^2)
\over q^2(q+k)^2} + 
g_i^r(k_0,\vec{k}\cdot\vec{q},\vec{q}^{\,2},\vec{k}^2)\right]
\frac{(q_0)^r}{(q_0^2-\mu^2)^r},\;\;\;
i=1,2,\dots , 14.
\ee
where $f_i^r$ and $g_i^r$ are polynomials in their arguments,
and the denominators $q_0^r$ have been replaced according to 
the prescription (\ref{presc}). This is the most general form of the
integrands from the diagrams with gluon or ghost loops. Notice that,
in particular, the ghost loop expression given by Eq. (\ref{ghloop})
yields, after projection, contributions of the kind given by the
$g_i^r$ terms above, which contains no on-shell poles.

The parameter $\mu$ regulates the originally ill defined sums and also
makes possible  the use of the formula (\ref{sumq0}), since now the
integrand is regular along the imaginary $q_0$ axis.
Hence (\ref{projpresc}) can be rewritten as
\be\label{projpresc1}
\Pi_{i}^{{\rm presc}}=\lim_{\mu\rightarrow
  0}\sum_{r=1}^{4}\int^{i\infty+\epsilon}_{-i\infty+\epsilon}
  {dq_0\over 2\pi i}\int d^{D-1}\vec{q}
\frac 1 2 \coth\left({q_0\over 2T}\right)\left\{
\frac{(q_0)^r}{(q_0^2-\mu^2)^r}\left[
{f_i^r(k_0,\vec{k}\cdot\vec{q},\vec{q}^{\,2},\vec{k}^2)
\over q^2(q+k)^2}
\right]+ q \leftrightarrow - q\right\} \, .
\ee
We have employed Eq. (\ref{Dreg}) and so the dimensionally
regularized integration of the $g_i^r$ terms has vanished. This important
property shows how the ghosts are effectively decoupled at finite
temperature.

Performing the $q_0$ integration by closing the integration
contour at right-hand side plane, we obtain from (\ref{projpresc1})
\be\label{eq3.7}
\Pi_{i}^{{\rm presc}}=\lim_{\mu\rightarrow  0}
\sum_{r=1}^{4}{d^{r-1} \over d\mu^{r-1}}\coth\left({\mu\over
      2T}\right)\int d^{D-1}\vec{q} \,\, 
G^r(\mu, k_0,\vec{k}\cdot\vec{q},\vec{q}^{\,2},\vec{k}^2).
\ee
In order to obtain the limit $\mu\rightarrow 0$, we need the following
expansions of the $\coth$ and its derivatives
\begin{eqnarray}\label{eq7.11}
&&\coth\left({\mu\over 2T}\right)={2T\over \mu}+{\cal O}(\mu)
\,,\quad \quad\quad
{d\over d\mu}\coth\left({\mu\over 2T}\right)=
-{2T\over \mu^2}+{1\over 6}{1\over T}+{\cal O}(\mu^2) \nnbb \\
&&{d^2\over  d\mu^2}\coth\left({\mu\over 2T}\right)=
{4T\over \mu^3}+{\cal O}(\mu)\,,\quad
{d^3\over  d\mu^3}\coth\left({\mu\over 2T}\right)=
-{12T\over \mu^4}-{1\over 60}{1\over T^3}+{\cal O}(\mu^2)\nonumber.
\end{eqnarray}
An important property of the contribution of the prescription poles is
that all the temperature dependence arises only from the expansions of
the hyperbolic cotangent shown above as {\it odd powers} of $T$.
Therefore, the results obtained in the previous section for the
leading $T^4$ and sub-leading $T^2$ and ${\log}(T)$ are not modified
by the temporal gauge prescription. 

It remains to be verified that the limit $\mu\rightarrow 0$ is well
defined. Our explicit calculations show that the results for all
projections do not involve inverse powers of $\mu$. Using the symmetry
of the angular integrals all the inverse powers of $\mu$ cancel 
and we obtain finite results given by
\begin{eqnarray}\label{eq4.5}
\Pi_{i}^{{\rm presc}}&=&T\int
d^{D-1}\vec{q}{F_i^1(k_0,\vec{k}\cdot\vec{q},\vec{q}^2,\vec{k}^2)\over
  \vec{q}^6
  [-k_0^2+(\vec{q}+\vec{k})^2]^5[-k_0^2+(\vec{q}-\vec{k})^2]^5}\nnbb
&+&{1\over T}\int
d^{D-1}\vec{q}{F_i^2(k_0,\vec{k}\cdot\vec{q},\vec{q}^2,\vec{k}^2)\over
  \vec{q}^4
  [-k_0^2+(\vec{q}+\vec{k})^2]^3[-k_0^2+(\vec{q}-\vec{k})^2]^3}\nnbb
&+&{1\over T^3}\int
d^{D-1}\vec{q}{F_i^3(k_0,\vec{k}\cdot\vec{q},\vec{q}^2,\vec{k}^2)\over
  \vec{q}^2
  [-k_0^2+(\vec{q}+\vec{k})^2][-k_0^2+(\vec{q}-\vec{k})^2]}. 
\end{eqnarray}
All these integrals are regular and can be done. In the Appendix D
we show in an explicit example a closed form result. For
$i=8,10,11,12$ we obtain $\Pi_{i}^{{\rm presc}}=0$, which is in
agreement with the  \'{}t Hooft identity given by Eq. (\ref{b5}).
We also show that  $\Pi_{i}^{{\rm presc}}=0$ for $i=7,9,13,14$.
Though the non-vanishing integrals ($i=1,2,3,4,5,6$) introduce an extra
temperature dependence, it is clear that they do not change the behavior of
the hard thermal loop expressions obtained in the previous section.

We remark that these non-vanishing integrals include both the thermal
and the zero temperature contributions (notice that
the integrands contains a $\coth$ instead of the purely thermal
part involving the Bose-Einstein distribution).
Had we computed the thermal part separately
we would be left with contributions which are divergent when
$\mu\rightarrow 0$ as well as the inverse powers of $T$. 
Since the dimensional regularization is employed only for the space part
of the integrals, the vacuum part also contains inverse powers of $\mu$
(only the fully dimensionally regularized zero temperature calculation is well
defined in the limit $\mu\rightarrow 0$ \cite{Leibbrandt:1994ki}).
It is remarkable that the inverse powers of $\mu$ in the
thermal part are exactly canceled by the corresponding ones in
the vacuum part and we are left only with the inverse powers of the
temperature. This property indicates that some of the ill-defined inverse
powers of $\mu$ have been replaced by a thermal regulated expression.
In order to understand why these prescription dependent parts are not
well defined when  $T\rightarrow 0$, one should notice that $\mu$
is a dimensionfull parameter which was made ``small'' in the sense
that $\mu << T$. Therefore, the prescription-dependent results
cannot be extended to the region $T\rightarrow 0$.

\section{Dispersion relations in a graviton plasma}

The sub-leading hard thermal loop contributions proportional to $T^2$
will produce modifications in the solution of
the dispersion relations describing the wave propagation in a graviton plasma.
The dispersion relations were carefully investigated in the case of
the leading $T^4$ contributions \cite{Rebhan:1991yr}. The
inclusion of sub-leading contributions has been considered in the
Feynman-de Donder gauge \cite{Brandt:1998hd}. Although the sub-leading 
modification of the solutions of the dispersion relations are
suppressed by a factor $G\, T^2\ll 1$, in relation to the order one part
arising from the $T^4$ contributions, one may be interested to know how the
gauge dependence of the graviton self-energy will affect these solutions
(the one-graviton function, which also contributes to the dispersion
relations, has no sub-leading
gauge dependent contributions at the one-loop order considered here).
In Yang-Mills theories, the problem of gauge-(in)dependence is well understood
since a theorem was proved by Kobes, Kunstatter and 
Rebhan (KKR) \cite{Kobes:1990xfKobes:1990dc}.
In an one-loop calculation, the gauge dependences of the location
of the poles of the gluon propagator are explained in terms of the
KKR identities. A well known example of this problem is the gauge
dependence of the plasmon damping constant (see \cite{Rebhan:2001wt}
for a recent review) and its solution by the Braaten and Pisarski resummation
scheme \cite{Braaten:1989kkBraaten:1989mz}.
As far as we know, a complete analysis of this problem, in the case of
gravity, is still missing. Therefore, we believe that it is important
to investigate how gauge dependent the graviton propagator is and
whether it is possible to extract gauge independent information.
In this regard, it is remarkable that the one-loop calculations of the 
QCD damping constant, in the axial gauge, though incomplete,
satisfy some of the necessary conditions required by any
physical quantity, being both gauge independent and
positive \cite{Kobes:1989cr}.

With this motivation, let us apply our axial gauge
results in the dispersion relations associated
with the {\it transverse traceless components} of the
{\it Jacobi equation} for small disturbances in the graviton
plasma \cite{Rebhan:1991yr}. Proceeding as in 
reference \cite{Brandt:1998hd}, the results given in Eq. (\ref{ct2}) 
(as well as the corresponding leading $T^4$ contributions) yields the
following dispersion relations for the three transverse traceless modes
\begin{eqnarray}\label{modABC}
&k^2&\left[1+{16\pi GT^2k^2 \over 15}\frac{\alpha}{n_0^2}\right]=16\pi
G\rho\left[\left({5\over9}+{1\over2}r^4L
    -{1\over6}r^2\right)\right.\nonumber \\
&&\left.+{5k^2\over\pi^2
    T^2}\left(r^2L+{5\over16}r^4L-{5\over48}r^2-{1\over12}+{1\over3}{1\over r^2} \right)\right]\nonumber
\\
&k^2&\left[1+{16\pi GT^2k^2\over 15}
\frac{\alpha}{n_0^2}
\left(1+{8\over3}{1\over r^2}\right)\right]=16\pi
G\rho\left[\left({2\over9}-2r^4L
    +{2\over3}r^2+{10\over9}{1\over r^2}\right)\right.\nonumber \\
&&\left.+{5k^2\over\pi^2 T^2}\left(-{13\over8}r^2L-{5\over4}r^4L+{5\over12}r^2+{7\over12}+{2\over3}{1\over r^2}\right)\right]\nonumber
\\
&k^2&\left[1+{16\pi GT^2k^2\over 15}
\frac{\alpha}{n_0^2}
\left(1+{8\over3}{1\over r^4}+{32\over9}{1\over r^2}\right)\right]=16\pi
G\rho\left[\left({8\over9}+3r^4L
    -r^2+{28\over27}{1\over r^2}\right)\right.\nonumber \\
&&\left.+{5k^2\over\pi^2 T^2}\left({5\over4}r^2L+{15\over8}r^4L-
{5\over8}r^2+{1\over24}+{1\over r^2}+
{4\over9}{1\over r^4}\right)\right];\;\;\; r^2\equiv{k^2\over \vec{k}^2}.
\end{eqnarray}
where $L(k)$ is given by Eq. (\ref{Llog}).

Let us now solve these relations in the region of real values 
of $k_0$ and $\vec k$, which is relevant for the propagation of waves,  
and then compare with the corresponding solutions previously obtained in the
Feynman-de Donder gauge. It is convenient to introduce the
dimensionless quantities $\bar k^2\equiv |\vec k|^2/(16\pi G\rho)$,
$\bar \omega^2\equiv\omega^2/(16\pi G\rho)$ 
and $\bar n_0^2\equiv n_0^2/(16\pi G\rho)$. We will also choose
$n_0^2=\omega^2$ so that the scale of the gauge fixing is compatible
with the momentum scale. For intermediate values of 
$\bar k$ and $\bar \omega$, the dispersion relations have to be
solved numerically and the results are qualitatively similar to
the ones shown in Fig. 3 of reference \cite{Brandt:1998hd}. In order
to discuss the specific issue of gauge dependence in terms
of well defined analytic expressions, we will consider the asymptotic regions
of very small and very large values of $\bar k,\,\bar\omega$.
In the limit $\bar k\rightarrow 0$ the solution of
the dispersion relations (\ref{modABC}) gives the following result
for the {\it plasma frequency} (this is the minimum frequency above
which propagating waves are supported by the plasma)
\be\label{wpx}
(\bar \omega_{\rm pl}^{axial})^2 = {\frac{22}{45}}
\left[1+\left(\frac{25}{6}-8\alpha\right)\frac{2\pi G T^2}{15}
\right],
\ee
where we have neglected higher powers of $G T^2$.
From Eqs. (4.15) of reference \cite{Brandt:1998hd} the same
limit $\bar k\rightarrow 0$ yields 
\be\label{wpF}
(\bar \omega_{\rm pl}^{cov.})^2 = {\frac{22}{45}}
\left[1-\left(\frac{14}{5}+(1-\xi)\right)\frac{32 \pi G T^2}{15}
\right],
\ee
were $\xi$ is the gauge parameter in the class of covariant gauges
and $\xi = 1$ defines the Feynman-de Donder gauge employed in
reference \cite{Brandt:1998hd}. An important
property of these results is that, in both classes of gauges,
there is the same strong dependence on the gauge parameter. In order to
understand this behavior, let us reintroduce the dimensionfull parameter
$16 \pi G \rho = (8/15)\,\pi^3\, G T^4$. Then, in both classes of gauges, one
can see that the gauge dependent subleading correction is of order
$(G T^4)(G T^2)$ which is of the same order as the 
{\it two-loop corrections}, not included in this calculation.
Therefore, the subleading contributions to the plasma frequency
constitute only a partial result at the one-loop order.

In the limit of high frequencies, $\bar\omega^2\sim\bar k^2 \gg 1$
the asymptotic behavior of the solutions is described by the
{\it thermal masses} 
$\bar m_{\rm I}^2\equiv \bar\omega_{\rm I}^2 -\bar k_{\rm I}^2$
(${\rm I=A,B,C}$). Using Eqs. (\ref{modABC}) it is straightforward to
obtain the following results
\be\label{mAx}
(\bar m_{\rm A}^{axial})^2 = \frac{5}{9}
\left[1 + \frac{8}{5}{\bar k^2\, \pi G T^2}\right],
\ee
\be\label{mBx}
(\bar m_{\rm B}^{axial})^2 = \frac{\sqrt{10}}{3}
\left[1 + \sqrt{10}\; \bar k \frac{4 \pi G T^2}{15}\right]\bar k
\ee
and
\be\label{mCx}
(\bar m_{\rm C}^{axial})^2 = \frac{2\sqrt{21}}{9}
\left[1 + \frac{4}{7} \bar k^2\, \pi G T^2 \right]\bar k.
\ee
When the same derivation is performed using Eqs. (4.15) of reference
\cite{Brandt:1998hd}, for arbitrary values of the gauge
parameter $\xi$, one obtains the following results
\be\label{mAF}
(\bar m_{\rm A}^{cov.})^2 = \frac{5}{9}
\left[1 - (9-\xi) \frac{32 \pi G T^2}{15}\right],
\ee
\be\label{mBF}
(\bar m_{\rm B}^{cov.})^2 = \frac{\sqrt{10}}{3}
\left[1 - (1-\xi) \frac{16 \pi G T^2}{15}\right]\bar k
\ee
and
\be\label{mCF}
(\bar m_{\rm C}^{cov.})^2 = \frac{2\sqrt{21}}{9}
\left[1 + \xi \frac{16 \pi G T^2}{15} \right]\bar k.
\ee

All these explicit examples clearly show the main differences
between these two distinct classes of gauges.
It is remarkable that the axial gauge subleading contributions
contain extra powers of $\bar k$ which makes then larger than
the corresponding corrections in the covariant gauges, of order
$(G T^4)(G T^2)$. Notice however, that the hard thermal loop condition,
$k^2\ll T^2$, implies that $\bar k^2\, G T^2 \ll 1$ so that the
subleading contributions will not exceed the leading ones.
As far as the gauge dependences are concerned, we remark that there
are no gauge parameter dependences in the axial gauge
results (for the choice $n_0^2=\omega^2\simeq |\vec k|^2$).
While this property is consistent with the necessary requirement
that any physical quantity should satisfy, the same is not true when
the masses $\bar m_{\rm I}$ (${\rm I=A,B,C}$) are computed in the
covariant gauges.

\section{Discussion}

In this article we have explicitly computed the thermal one- and two-graviton
functions in the temporal gauge. We have applied 
Leibbrandt's \cite{Leibbrandt:1994ki} prescription to deal with
the temporal gauge poles at finite temperature.
This calculation provides a rather non-trivial explicit
verification of the gauge
invariance properties of the hard thermal loop contributions. Indeed,
the leading $T^4$ behavior is in agreement with previous calculations
in covariant gauges. The subleading contributions of order $T^2$ have
a gauge dependence in agreement with the  \'{}t Hooft identities. 
We have also compared our ${\log} T$ contributions with
the residue of the ultraviolet pole of the dimensionally
regularized zero temperature graviton self energy,
given in Ref. \cite{Capper:1982rc}, and found that they are the same
(this property has also been verified in the Feynman-de Donder 
gauge \cite{Brandt:1998hd}). Our results include also the full gauge
parameter dependence, as shown in the Appendix C.

Our explicit calculation indicates that the temporal gauge may be
consistently employed even in the highly non-trivial case of thermal
gravity. The form of the prescription poles in Eq. (\ref{presc}) do
not change the hard thermal loop behavior of our main result given by
Eq. (\ref{gravForw}). An important property of this forward scattering
amplitude is that, as opposite to the covariant gauges,
it does not involve {\it thermal ghosts} and the gauge parameter
dependence is linear in $\alpha$.

In the analysis of the dispersion relations we have included the hard
thermal loop subleading contributions proportional to $T^2$ and
compared the structure of the gauge dependence with similar
calculations which were performed earlier in the Feynman gauge.
As expected from general formal arguments there are gauge dependent
contributions which arises from the subleading $T^2$ terms in the
graviton self-energy. By power counting, some of the gauge dependent
terms are of the same order as the two-loop contributions. 
However, the subleading terms, when computed in the axial gauge,
are such that their contributions to the asymptotic masses are
enhanced by extra powers of $\bar k$ and have a weaker gauge dependence.
This behavior is analogous to what happens in QCD, where the 
plasmon damping constant (which also is subleading in the temperature)
has a weaker gauge dependence when computed in non-covariant gauges.

\begin{acknowledgments}
F.T.B. would like to thank CNPq for financial support and
Prof. J. Frenkel for many helpful discussions.
F.R.M and B.C-M would like to thank Fapesp for providing the financial
support for their graduate program.
\end{acknowledgments}

\appendix

\section{Ward identities}\label{apa}
In this appendix we derive the identities which must be satisfied by
the vertex functions generated from an action which is invariant under
coordinate transformations. These identities provide an important
consistency check of the gravitational Feynman rules as well as the
leading high temperature thermal Green's functions. 

The invariance of an action $S$ can be expressed as follows
\be
\label{wardgen}
\delta\, S = \int \text{d}^4x\,
\frac{\delta {\cal L}(x)}{\delta \phi_{\mu\nu}(x)}\,
\delta \phi_{\mu\nu}(x)=0.
\ee
Let us choose the following coordinate transformation with an infinitesimal
parameter $\de_\mu(x)$
\[
{x^{\prime}}^\mu=x^\mu +\de^\mu \left(x\right).
\] 
Performing the transformation in the metric
\begin{eqnarray}
g^\prime_{\mu\nu}(x^\prime) & = & 
\frac{\partial x^\alpha}{\partial {x^\prime}^\mu}
\frac{\partial x^\beta }{\partial {x^\prime}^\nu} \,g_{\alpha\beta}
= g_{\mu\nu}(x) 
-g_{\alpha\nu}(x)\,\partial_\mu\,\de^\alpha
-g_{\alpha\mu}(x)\,\partial_\nu\,\de^\alpha
\nnbb
& = &  g^\prime_{\mu\nu}(x) + 
\de^\lambda\,\partial_\lambda\, g^\prime_{\mu\nu}(x),
\end{eqnarray}
we obtain
\begin{eqnarray}
\label{delphi}
g^\prime_{\mu\nu}(x) - g_{\mu\nu}(x)\equiv
\delta g_{\mu \nu }& = & \kappa\,\delta \phi_{\mu \nu }=
  - g_{\mu \sigma}\,\partial_\nu \de^\sigma
  - g_{\nu \sigma}\,\partial_\mu \de^\sigma
-\de_\lambda\left(\partial^\lambda g_{\mu \nu }\right) \nnbb
& = & -\partial_\nu\,\de_\mu 
     - \partial_\mu\,\de_\nu -
\kappa\left[\phi_{\mu\sigma}\,(\partial_\nu\de^\sigma) +
            \phi_{\nu\sigma}\,(\partial_\mu\de^\sigma) +
            \de^\lambda\,(\partial_\lambda\phi_{\mu\nu})\right],
\end{eqnarray}
where we have used Eq. (\ref{gravdef}).
Inserting (\ref{delphi}) into (\ref{wardgen}) and using integration
by parts, we obtain
\be\label{ward1}
\int \text{d}^4 x\, \de^\lambda
\left(\eta_{\nu\lambda}\, \partial_\mu +
      \eta_{\mu\lambda}\, \partial_\nu \right)
\frac{\delta {\cal L}}{\delta \phi_{\mu \nu }} = - \kappa
\int \text{d}^4 x\,\de^\lambda\left( 
\partial_\nu\,\phi_{\mu\lambda} + 
\partial_\mu\,\phi_{\nu\lambda} + 
(\partial_\lambda\,\phi_{\mu\nu})\right)
\frac{\delta {\cal L}}{\delta \phi_{\mu \nu }}
\ee
Taking functional derivatives of Eq. (\ref{ward1}) one obtains
the following Ward identities in momentum space
\begin{eqnarray}\label{v2v1}
\frac 1 \kappa \chi^0_{\mu\nu\lambda}(k_1)V^{2\,\mu\nu}_{\alpha\beta}(k_1,k_2) & = & 
- \chi^1_{\mu\nu\alpha\beta\lambda}(k_1,k_2)\,V^{1\,\mu\nu}(k_1+k_2=0)
\\ \nnbb \label{v3v2}
\frac 1 \kappa \chi^0_{\mu\nu\lambda}(k_1)
V^{3\,\mu\nu}_{\,\,\,\,\,\,\,\,\alpha\beta\delta\gamma}(k_1,k_2,k_3)
& = & 
-\chi^1_{\mu\nu\alpha\beta\lambda}(k_1,k_2)
V^{2\,\mu\nu}_{\,\,\,\,\,\,\,\,\delta\gamma}(k_1+k_2,k_3) 
 \nnbb & &
-\chi^1_{\mu\nu\delta\gamma\lambda}(k_1,k_3)
V^{2\,\mu\nu}_{\,\,\,\,\,\,\,\,\alpha\beta}(k_1+k_3,k_2) 
\\ \nnbb \label{v4v3}
\frac 1 \kappa \chi^0_{\mu\nu\lambda}(k_1)
V^{4\,\mu\nu}_{\,\,\,\,\,\,\,\,\alpha\beta\delta\gamma\tau\sigma}
(k_1,k_2,k_3,k_4)
& = & 
- \chi^1_{\mu\nu\alpha\beta\lambda}(k_1,k_2)
V^{3\,\mu\nu}_{\,\,\,\,\,\,\,\,\delta\gamma\tau\sigma}
(k_1+k_2,k_3,k_4)  \nnbb & &
- \chi^1_{\mu\nu\delta\gamma\lambda}(k_1,k_3)
V^{3\,\mu\nu}_{\,\,\,\,\,\,\,\,\alpha\beta\tau\sigma}
(k_1+k_3,k_2,k_4)   \nnbb & &
-\chi^1_{\mu\nu\tau\sigma\lambda}(k_1,k_4)
V^{3\,\mu\nu}_{\,\,\,\,\,\,\,\,\alpha\beta\delta\gamma}
(k_1+k_4,k_2,k_3),
\end{eqnarray}
where
\begin{eqnarray}\label{chi0}
\chi^0_{\mu\nu\lambda}(k_1)&=&{k_1}_\mu\eta_{\nu\lambda}+
                              {k_1}_\nu\eta_{\mu\lambda}
\\\label{chi1}
\chi^1_{\mu\nu\alpha\beta\lambda}(k_1,k_2)&=&
{k_1}_\nu\eta_{\alpha\mu}\eta_{\lambda\beta}+
{k_1}_\mu\eta_{\alpha\lambda}\eta_{\nu\beta}+
{k_2}_\lambda\eta_{\alpha\mu}\eta_{\beta\nu},
\end{eqnarray}
and the vertices $V^n$ are the momentum space expressions for the 
$n$-th functional derivatives computed at $\phi_{\mu\nu}=0$ 
(momentum conservation is understood in all identities).

In the case of a tree-level action there is no one-point 
``vertex'' $V^{1\,\mu\nu}$ and so
the quadratic term satisfies the transversality condition
\be\label{transv1}
\chi^0_{\mu\nu\lambda}(k_1)V^{2\,\mu\nu}_{\alpha\beta}(k_1,k_2)=0.
\ee
This may not be the case for an effective gauge invariant Lagrangian.
Indeed, it is well know that the one-graviton function is non-zero at
finite temperature.

\section{Gravitational {\rm \'{}t} Hooft identities}
\label{apb}
The imaginary time formalism at finite temperature follows closely 
the corresponding formalism at $T=0$. Consequently, the \'{}t Hooft
identities at finite $T$ would be identical to the ones at $T=0$,
if there were no 1-particle tadpole contributions
(such terms vanish at $T=0$ in the dimensional regularization scheme).
However, since the tadpole is exactly proportional to $T^4$, it will
not affect the identities involving the sub-leading contributions.
To derive these, we start from the action
\be
  \label{Iaction}
I =  \int {\rm d}^4 x  {\rm d}^4 y \phi_{\mu\nu}(x) S_{{\rm sub}}^{\mu\nu\,\alpha\beta}(x-y)
                              \phi_{\alpha\beta}(y) +
  \int {\rm d}^4 x  {\rm d}^4 y J^{\mu\nu}(x) X_{\mu\nu\, \lambda}(x-y)
                              \eta^{\lambda}(y) + \cdots.
\ee
Here $S_{{\rm sub}}^{\mu\nu\,\alpha\beta}$ denotes the tree order quadratic
term plus the sub-leading contributions to the graviton 2-point function 
and $X_{\mu\nu\, \lambda}$ represents
the tensor generated by a gauge transformation of the graviton field which
is given to lowest order, in the momentum space,
by Eq. (\ref{chi0}). $J^{\mu\nu}$ is an
external source, $\eta^{\lambda}$ represents the ghost field and $\cdots$ stand
for terms which are not relevant for our purpose. The \'{}t Hooft identity
involving the graviton self-energy function is a consequence of the BRST
invariance of the action $I$:
\be
\label{brst}
\int{\rm d}^4 x\frac{\delta I}{\delta J^{\mu\nu}(x)} 
              \frac{\delta I}{\delta \phi_{\mu\nu}(x)} = 0
\ee
In general, Eq. (\ref{brst}) implies the \'{}t Hooft identity
\be
\label{gentHooft1}
X_{\mu\nu\; \lambda}\, S_{{\rm sub}}^{\mu\nu\,\alpha\beta} = 0 ,
\ee
which can be written to second order as
\be
\label{gentHooft2}
X^{(0)}_{\mu\nu\,\lambda}\, \Pi_{{\rm sub}}^{\mu\nu\,\alpha\beta} = 
- X^{(1)}_{\mu\nu\,\lambda}\, V^{2\; \mu\nu\,\alpha\beta},
\ee
where $V^{2\; \mu\nu\,\alpha\beta}$ satisfies the
identity (\ref{transv1}). Using Eq. (\ref{transv1})
we see that (\ref{gentHooft2}) leads immediately to the
\'{}t Hooft identity 
\be\label{b5}
\chi^{(0)}_{\mu\nu\lambda}(k)
\Pi^{\mu\nu\alpha\beta}_{{\rm sub}}(k,u)\chi^{(0)}_{\alpha\beta
 \delta}(k)=0 \, .
\ee
It is straightforward to show that the identity (\ref{b5}) implies
that ${\Pi_{{\rm sub}}}_{8}={\Pi_{{\rm sub}}}_{10}
={\Pi_{{\rm sub}}}_{11}={\Pi_{{\rm sub}}}_{12}=0$ and so these projections
have a temperature behavior which is at most proportional to $T^4$.

In order to verify (\ref{hooft}) we need to calculate the tensor
$\chi^{(1)}_{\mu\nu\lambda}$ which appears in (\ref{gentHooft2}). In
this way we need the source-graviton-ghost vertex which can be
obtained from the Lagrangian \cite{Delbourgo:1985wz} 
\be
{\cal L}_S= \kappa J^{\mu\nu}D_{\mu\nu\lambda}\epsilon^\lambda
\ee 
Using the transformation $\delta g_{\mu\nu}=\kappa
D_{\mu\nu\lambda}\epsilon^\lambda$ we obtain
\be\label{eqc.3}
\begin{array}{c}
  \hbox{\epsfxsize=.15\textwidth
\epsfbox{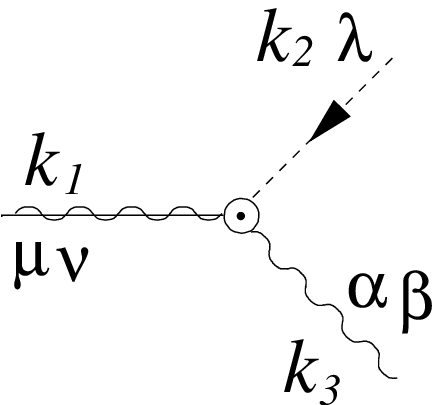}}
\end{array}=-i 
{\kappa \over 2}
\left[\eta_{\alpha\lambda}\eta_{\beta\nu}{k_2}_\mu+
      \eta_{\alpha\mu}\eta_{\beta\lambda}{k_2}_\nu+
      \eta_{\alpha\mu}\eta_{\beta\nu}{k_3}_\lambda\right]
  + \alpha\leftrightarrow \beta
\ee

The diagram in Fig. 3 can be now calculated using vertex
(\ref{eqc.3}) and Feynman rules (\ref{gravprop}), (\ref{ghprop}) and
(\ref{Ggg}). Expanding $\chi^{(1)}_{\mu\nu\lambda}$ in the base
shown in Table II
\begin{table}[htbp]\begin{center}
\begin{tabular}{c}\hline
$k_\mu k_\nu k_\lambda$ \\
$u_\mu u_\nu u_\lambda$ \\
$k_\lambda u_\mu u_\nu$ \\
$u_\lambda k_\mu k_\nu$\\
$k_\mu u_\nu u_\lambda+k_\nu u_\mu u_\lambda$ \\
$u_\mu k_\nu k_\lambda+u_\nu k_\mu k_\lambda$ \\ 
$\eta_{\mu\nu}k_\lambda$ \\
$\eta_{\mu\nu}u_\lambda$ \\
$k_{\mu}\eta_{\nu\lambda}+k_{\nu}\eta_{\mu\lambda}$ \\
$u_{\mu}\eta_{\nu\lambda}+u_{\nu}\eta_{\mu\lambda}$\\
\hline
\end{tabular}\caption{10 independent tensors base}\end{center}
\end{table}
and using the forward scattering method as we did for the one- and
two-graviton functions, we obtain the following {\it leading} $T^2$
contribution for $\chi^{(1)}_{\mu\nu\lambda}$ 
\be\label{eqc.4}
\chi^{(1)}_{\mu\nu\lambda}={T^2\over
  18}k_0\left[\eta_{\lambda\nu}u_\mu+\eta_{\lambda\mu}u_\nu-2u_\mu
  u_\nu u_\lambda \right]
\ee
Contracting (\ref{eqc.4}) with $V^{2\,\mu\nu\alpha\beta}$ yields
\begin{eqnarray}\label{eqc.5}
\chi^{(1)}_{\mu\nu\lambda}V^{2\,\mu\nu\alpha\beta}&=&{T^2\over  18}k_0
\left\{2k\cdot u \left(k_\alpha u_\beta+k_\beta u_\alpha\right)
  +k\cdot
  u\left(2\eta_{\alpha\beta}k_\lambda-\eta_{\lambda\beta}k_\alpha-
    \eta_{\lambda\alpha}k_\beta\right)\right.\nnbb \\
&+&\left.{k^2}\left(\eta_{\alpha\lambda}u_\beta+
                    \eta_{\beta\lambda}u_\alpha-2u_\alpha
    u_\beta u_\lambda\right) - k_\lambda\left(k_\alpha u_\beta
    +k_\beta u_\alpha\right) -2k\cdot
  u^2\eta_{\alpha\beta}u_\lambda\right\} \nonumber
\end{eqnarray}
Using the result for $\Pi_{T^2}^{\mu\nu\alpha\beta}$, which can be
obtained inserting Eq. (\ref{ct2}) into (\ref{selfgen}), we have verified that
the contraction with  $\chi^{(0)}_{\mu\nu\lambda}$ yields Eq. (\ref{eqc.5})
with opposite sign in complete agreement with Eq. (\ref{hooft}).

\section{The $\log T$ contributions for arbitrary values of $\alpha$}

In this appendix we complement the result presented in Eq. (\ref{logT})
and include the contributions proportional to the gauge parameter $\alpha$.
We have obtained the following results for the 14 projections 
(see Eq. (\ref{proj}))
\begin{eqnarray}
\Pi_1^{\log} &=& {\kappa^2\, k^4\over 15 \pi^2}\log {(T)}\left[\frac{1}{32}(1312 y^2 - 304
  y -363)-{1\over 28}\frac{\alpha}{n_0^2}k^2 (4336
  y^3-1948y^2-1272y+39)\right] \nonumber \\
\Pi_2^{\log} &=&  - {\kappa^2\, k^4\over 15\pi^2}  (y-1)\log {(T)}\left[\frac{1}{16}(8y -113)
  -{1\over 14}\frac{\alpha}{n_0^2}k^2(1332y^2-928y-19)\right] \nonumber \\
\Pi_3^{\log} &=& {\kappa^2\, k^4\over 15\pi^2}  (y-1)^2
  \log {(T)}\left[\frac{17}{16}-\frac{1}{28}\frac{\alpha}{n_0^2}k^2 (76y+1)\right] \nonumber \\
\Pi_4^{\log} &=& {\kappa^2\, k^4\over 15\pi^2}
  \log {(T)}\left[\frac{1}{32}(352 y^2 -464 y -23
  )-\frac{1}{7}\frac{\alpha}{n_0^2}k^2 (y-1)(1084y^2-243y-1)\right] \nonumber \\
\Pi_5^{\log} &=&  - {\kappa^2\, k^4\over 15\pi^2}
  (y-1)\log {(T)}\left[\frac{1}{16}(56 y+59)+\frac{1}{7}\frac{\alpha}{n_0^2}k^2
  (y-1)(174y+1) \right]
\nonumber \\
\Pi_6^{\log} &=&  {4\,\kappa^2  k^4\over 3\pi^2}\log {(T)} y (y-1) \nonumber \\
\Pi_7^{\log} &=&  
\Pi_8^{\log} =    
\Pi_9^{\log} =    
\Pi_{10}^{\log} = 
\Pi_{11}^{\log} = 
\Pi_{12}^{\log} =  0 \nonumber \\
\Pi_{13}^{\log} &=& {2\kappa^2  k^4\over 3\pi^2}\log {(T)}
y^2 (y-1)\nonumber \\
\Pi_{14}^{\log} &=&  0 ,
\end{eqnarray}
where we are using the quantity $y\equiv{{k_0^2}\over {k ^2}}$ in
order to compare with the zero temperature results of reference 
\cite{Capper:1982rc}.

In terms of these projections, the coefficients of the 
transverse traceless components of the graviton self-energy,
as defined for instance in reference \cite{Rebhan:1991yr},
can be written as
\begin{eqnarray}
c_{\rm A} & = & {\kappa^2\, k^4\over 15 \pi^2} \log {(T)}
\left[\frac 5 2 y^2 + 2 y + \frac{19}{64} - 
\alpha\frac{k_0^2}{n_0^2}\left(\frac{24}{7} +\frac{39}{56}\frac 1 y
\right)\right]
\nonumber \\
c_{\rm B} & = & {\kappa^2\, k^4\over 15 \pi^2} \log {(T)}
\left[5 y^2 + \frac 1 8 y - \frac{181}{64} +
\alpha\frac{k_0^2}{n_0^2}\left(\frac{135}{7} +\frac{3}{8}\frac 1 y
-\frac{333 y}{14}\right)\right] 
\nonumber \\
c_{\rm C} & = & {\kappa^2\, k^4\over 15 \pi^2} \log {(T)}
\left[\frac{11}{6} y^2 -\frac{25}{6} y - \frac{71}{192} + 
\alpha\frac{k_0^2}{n_0^2}\left(\frac{53}{21} - \frac{168}{56}\frac 1 y
+\frac{115}{6} y - \frac{542}{21} y^2 \right)\right].
\end{eqnarray}
These expressions show that the dispersion relations associated with
the transverse traceless modes will, in general, be gauge dependent
at this order of perturbation theory.

\section{}

As an example of the calculation shown in subsection 4.2 we will
calculate a contribution of the prescription poles for the projection
$\Pi^{{\rm presc}}_6$. This contribution can be written as
\be
\Pi_{6}^{{\rm presc}}=\frac{\kappa^2}{k_0}\,T\,\sum_{q_0}\int
\frac{d^{D-1}\vec{q}}{(2\pi)^{D-1}}\frac{1}{2}\frac{(D-3)\,D}{D-2}
\left[{\vec{q}^{\,4}+(5-2D)\,k_0^4+(2D-6)\,k_0^2\,(\vec{k}^2+\vec{q}^{\,2})+\vec{k}^4
-2\vec{q}^{\,2} \vec{k}^2 \over q_0\;(q+k)^2}\right] .
\ee 
Using the prescription (\ref{presc}) and Eq. (\ref{sumq0}) we obtain
\begin{eqnarray}\small
\hspace*{-2.5 cm}\Pi_{6}^{{\rm presc}}&=&\frac{\kappa^2}{k_0}\,
\lim_{\mu\rightarrow 0}\int^{i\infty+\epsilon}_{-i\infty+\epsilon}{dq_0\over
  4\pi i}\int
{d^{D-1}\vec{q}\over (2\pi i)^{D-1}}\coth\left({q_0\over
      2T}\right)\frac{D}{2}\frac{(D-3)}{D-2}\nnbb
& & \left\{{q_0\left[ \vec{q}^{\,4}+(5-2D)\,k_0^4+(2D-6)\,k_0^2
\,(\vec{k}^2+\vec{q}^{\,2})+\vec{k}^4
-2\vec{q}^{\,2} \vec{k}^2\right] \over (q_0^2-\mu^2)\;(q+k)^2}
+ q \leftrightarrow -q  {{~}\atop{~}}\right\} .
\end{eqnarray}\normalsize
Closing the integration contour at right-hand side plane and 
expanding in terms of a power series in $\mu$ we obtain
\begin{eqnarray}
\Pi_{6}^{{\rm presc}}&=& \kappa^2\,T\,\int
{d^{D-1}\vec{q}\over (2\pi i)^{D-1}}\frac{D}{2}\frac{(D-3)}{D-2}
\left\{{(2D-6)\,k_0^2\,(\vec{q}^{\,2}+\vec{k}^2)+(5-2D)\,k_0^4+(\vec{k}^2-\vec{q}^{\,2})^2 \over
  \left[k_0^2-(\vec{q}+\vec{k})^2\right]^2} +\vec{k} \leftrightarrow
-\vec{k}\right\}  \nnbb \label{eqc3}
&=& \kappa^2\,T\,\int
{d^{D-1}\vec{q}\over (2\pi i)^{D-1}}\frac{D(D-3)}{D-2}\,  \left[
  {-(2D-6)\,k_4^2\,(\vec{q}^{\,2}+2\vec{k}^2)+(5-2D)\,k_4^4+\vec{q}^{\,4}
+4\left(\vec{k}\cdot \vec{q}\right)^2 \over (k_4^2+\vec{q}^2)^2}\right]
\end{eqnarray}
where $k_4=ik_0$ and we have performed a shift
$\vec q\rightarrow \vec q - \vec k$. In the limit $D\rightarrow 4$ we obtain
\be
\Pi_{6}^{{\rm presc}}={2 \kappa^2\,T\over \pi}|k_4|(|k_4|^2-\vec{k}^2)
\ee

The contributions to the projections
$\Pi^{{\rm presc}}_7$, $\Pi^{{\rm presc}}_9$, $\Pi^{{\rm presc}}_{13}$ and
$\Pi^{{\rm presc}}_{14}$ are obtained in a similar way and we find
that they vanish. However, for the projections 
$\Pi^{{\rm presc}}_1$, $\Pi^{{\rm presc}}_2$, $\Pi^{{\rm presc}}_3$,
$\Pi^{{\rm presc}}_4$ and  $\Pi^{{\rm presc}}_5$ we have more involved 
expressions containing inverse powers of $T$ as in Eq. (\ref{eq4.5}).

\end{document}